\documentclass[10pt,journal]{IEEEtran}
\usepackage{amsmath,amsfonts}
\usepackage{algorithmic}
\usepackage{algorithm}
\usepackage{array}
\usepackage[caption=false,font=normalsize,labelfont=sf,textfont=sf]{subfig}
\usepackage{textcomp}
\usepackage{stfloats}
\usepackage{url}
\usepackage{verbatim}
\usepackage{graphicx}
\usepackage{cite}
\usepackage{hyperref}
\pdfoutput=1
\newtheorem{theorem}{Theorem}
\newtheorem{definition}[theorem]{Definition}

\newtheorem{corollary}[theorem]{Corollary}

\begin{document}

\title{Theoretical Bounds for the Size of Elementary Trapping Sets by Graphic Methods}

\author{
  \IEEEauthorblockN{Haoran Xiong\IEEEauthorrefmark{2}\IEEEauthorrefmark{3}, 
   Zicheng Ye\IEEEauthorrefmark{2}\IEEEauthorrefmark{3}, 
   Huazi Zhang\IEEEauthorrefmark{4}, 
   Jun Wang\IEEEauthorrefmark{4}, 
   Ke Liu\IEEEauthorrefmark{4}, 
   Dawei Yin\IEEEauthorrefmark{5},\\
   Guanghui Wang\IEEEauthorrefmark{5},
   Guiying Yan\IEEEauthorrefmark{2}\IEEEauthorrefmark{3}, 
   and Zhiming Ma\IEEEauthorrefmark{2}\IEEEauthorrefmark{3}}

   \IEEEauthorblockA{\IEEEauthorrefmark{2}
                     University of Chinese Academy of Sciences}

   \IEEEauthorblockA{\IEEEauthorrefmark{3}
                     Academy of Mathematics and Systems Science, CAS }

   \IEEEauthorblockA{\IEEEauthorrefmark{4}
                     Huawei Technologies Co. Ltd.}

   \IEEEauthorblockA{\IEEEauthorrefmark{5}
                     School of Mathematics, Shandong University}

    Email: \{xionghaoran, yezicheng\}@amss.ac.cn, \{zhanghuazi, justin.wangjun, liuke79\}@huawei.com,\\
    daweiyin@mail.sdu.edu.cn, ghwang@sdu.edu.cn, yangy@amss.ac.cn, mazm@amt.ac.cn
    }




\maketitle

\begin{abstract}
  Elementary trapping sets (ETSs) are the main culprits for the performance of LDPC codes in the error floor region. Due to the large quantity, complex structures, and computational difficulties of ETSs, how to eliminate dominant ETSs in designing LDPC codes becomes a pivotal issue to improve the error floor behavior. In practice, researchers commonly address this problem by avoiding some special graph structures to free specific ETSs in Tanner graph. In this paper, we deduce the accurate Tur{\text{\'a}}n number of $\theta(1,2,2)$ and prove that all $(a,b)$-ETSs in Tanner graph with variable-regular degree $d_L(v)=\gamma$ must satisfy the bound $b\geq a\gamma-\frac{1}{2}a^2$, which improves the lower bound obtained by Amirzade when the girth is 6. For the case of girth 8, by limiting the relation between any two 8-cycles in the Tanner graph, we prove a similar inequality $b\geq a\gamma-\frac{a(\sqrt{8a-7}-1)}{2}$. The simulation results show that the designed codes have good performance with lower error floor over additive white Gaussian noise channels.
\end{abstract}


\section{Introduction}
\IEEEPARstart{L}{ow-density} parity-check (LDPC) codes are capacity-approaching codes, which are widely used in Wi-Fi, optical microwave, storage and other systems, and are adopted as 5G eMBB data channel coding \cite{gallager1962low, chung2001design}. Quasi-cyclic LDPC (QC-LDPC) codes hold significant importance due to their excellent error correction capabilities and efficient hardware coding implementation \cite{townsend1967self, okamura2003designing}. However, there is a problematic phenomenon called error floor under iterative decoding of LDPC codes, which is characterized by a slow decrease of error rate curves as channel quality improves.

The error floor behavior of LDPC codes is mainly caused by the graphical structures of the code's Tanner graph, which is known as trapping sets \cite{richardson2003error}. An $(a,b)$ trapping set is an induced subgraph with $a$ variable nodes, $b$ check nodes of odd degree and an arbitrary number of even degree check nodes in the Tanner graph. An elementary trapping set (ETS) is a trapping set whose all check nodes are of degree 1 or 2, which are the most harmful ones among trapping sets \cite{cole2006general, milenkovic2006asymptotic}.

However, the number of non-isomorphic structures of ETSs with different values of $a$ and $b$ is hard to count. Indeed, McGregor \emph{at el.} \cite{mcgregor2010hardness} proved the problem of finding the minimum size of trapping sets in Tanner graph of a LDPC code is NP-hard. For simplicity, we focus on ETSs of variable-regular Tanner graphs in the following parts, where we can use a special graph known as variable node (VN) graph \cite{koetter2004pseudo} to simplify an ETS. What's more, there is a one-to-one correspondence between an ETS and its VN graph if the Tanner graph is variable-regular. 

Due to the diversity of ETSs, many researchers tried to place restrictions on Tanner graph to remove small ETSs. X. Tao \emph{et al.} \cite{tao2017construction} constructed QC-LDPC codes with variable-regular degree $d_L(v)=3$ and girth $g=8$ which are free of all $(a,b)$-ETSs with $a\leq8$ and $b\leq3$ by eliminating some certain 8-cycles. For $g=8$ and $d_L(v)=3$ or 4, Naseri \emph{et al.} \cite{naseri2019construction} proved that if all 8-cycles generated by two different rows are avoided and all 8-cycles generated by three different rows from the base matrix are controlled, a large range of ETSs can be eliminated. For $g=8$ and different variable-degree $d_L(v)\in \{4,5,6\}$, using edge-coloring technique, Amirzade \emph{et al.} \cite{amirzade2021qc} proved that if all 8-cycles are generated by four different rows from the base matrix, then several small ETSs can be free of. In \cite{hashemi2015characterization, karimi2014characterization}, Amir H. Banihashemi \emph{et al.} proved by computer programming that any ETS with relatively small $a$ and $b$ is generated by a short cycle or some non-cycle graphs whose basic structures are theta graphs and dumbbell graphs. By avoiding 8-cycle with a chord that is exactly theta graph $\theta(1,2,2)$ in VN graph, Amirzade \emph{et al.} \cite{amirzade2022protograph} constructed QC-LDPC codes with girth $g=6$ which are free of all $(a,b)$-ETSs with $a\leq5$ and $b\leq3$ for $d_L(v)=3$ and $a\leq7$ and $b\leq4$ for $d_L(v)=4$. In more detail, they deduced the bound $b\geq a\gamma-\frac{2a^3}{4a-3}$ for $(a,b)$-ETSs which serves as an upper bound of the Tur\'an number of $\theta(1,2,2)$. The Tur{\'a}n problem is the most typical problem in extremal graph theory, which mainly studies the maximum number of edges (called Tur{\'a}n number) in a given graph (or hypergraph) without some special substructures.

In this paper, we prove that if we free the theta graph $\theta(1,2,2)$ in the VN graph when the girth $g=6$, then several small ETSs can be eliminated. Indeed, by determining the accurate value of Tur{\text{\'a}}n number for theta graph $\theta(1,2,2)$, we prove that all $(a,b)$-ETSs in Tanner graph with variable-regular degree $d_L(v)=\gamma$ must satisfy the bound $b\geq a\gamma-\frac{1}{2}a^2$, which improves the bound $b\geq a\gamma-\frac{2a^3}{4a-3}$ in \cite{amirzade2022protograph}. Also, we notice that the minimum $a$ calculated by $b\geq a\gamma-\frac{1}{2}a^2$ is coincide with the values obtained by enumeration. Moreover, we consider eliminating theta graph $\theta(2,2,2)$ for the case girth $g=8$ and prove that all $(a,b)$-ETSs are free of with $(a,b)$ not satisfying the bound $b\geq a\gamma-\frac{a(\sqrt{8a-7}-1)}{2}$. The numerical results are presented to demonstrate the effectiveness of eliminating theta graph $\theta(2,2,2)$, when the girth is 8.

The rest of this paper is organized as follows. In section 2, the basic definitions and notations are presented. We determine Tur{\text{\'a}}n number of special theta graph in section 3. Indeed, we obtain the accurate value of $ex(n,\theta(1,2,2))$ by induction and an upper bound of $ex(n,\theta(2,2,2))$ by proving some properties of its extremal graph. In section 4, the results in section 3 are applied to coding theory and we get bounds for the parameters $a,b$ and $\gamma$ while concerning an $(a,b)$-ETS in a variable-regular Tanner graph with $d_L(v)=\gamma$. By these bounds, several small ETSs can be free of. Section 5 presents construction examples based on our methods and the corresponding numerical results. The paper is concluded in section 6.

\section{Preliminaries}
A graph $G$ is a pair of sets $(V,E)$ where $V$ is a nonempty set of objects and $E$ is a (possibly empty) set of unordered pairs of elements of $V$. The elements of $V$ are called the vertices of $G$ and the elements of $E$ are called the edges of $G$. For $u,v\in V$, we say $u$ is adjacent to $v$ or $u$ and $v$ are neighbors if there is an edge between $u$ and $v$. The set of neighbors of a vertex $v$ is called the neighborhood of $v$, denoted by $N(v)=\{u\in V|(u,v)\in E\}$. If a graph $G$ contains two vertices which are incident to more than one edge, then the graph is called a multigraph. For an edge that its endpoints are the same, we call it a loop. A simple graph is the graph has no multiple edges or loops. A simple graph with $n$ vertices is complete if every two of its vertices are adjacent and we denote this graph by $K_n.$ A simple graph is bipartite if its vertices can be partitioned into two sets $V_1$, $V_2$ that no edge joins two vertices in the same set. A bipartite graph is complete if for any $v\in V_1$ and $u\in V_2$, then the edge $(u,v)\in E$. We denote a complete bipartite graph by $K_{m,n}$ if $|V_1|=m$, $|V_2|=n.$

There is a bipartite graph $G=(L \cup R,E)$ constructed from the check matrix $H$ of an LDPC code, which is known as Tanner graph \cite{tanner1981recursive}. $L$ labels all variable nodes corresponding to the columns of $H$ and $R$ labels all check nodes corresponding to the rows. The $i$-th check node is adjacent with the $j$-th variable node if and only if $H_{ij}=1$. We denote the degree of vertex $v$ in $G$ by $d_G(v)$, which is the number of its adjacent vertices. The bipartite graph is variable-regular if $d_L(v)=\gamma\geq1$ for all $v\in L$.

For a fixed positive integer $p$ referred to as lifting degree, a $(\gamma,\eta)$-regular QC-LDPC code of length $N=p\eta$ is an LDPC code whose column weight and row weight are $\gamma$ and $\eta$, respectively. The parity check matrix $H$ can be represented as \cite{fossorier2004quasicyclic}:
\begin{equation}
    \label{eq1}
    H = 
\begin{bmatrix}
    I(0) & I(0) & \dots & I(0)\\
    I(0) & I(p_{1,1}) & \dots & I(p_{1,\eta-1})\\
    \vdots & \vdots & \ddots & \vdots\\
    I(0) & I(p_{\gamma-1,1}) & \dots & I(p_{\gamma-1,\eta-1})\\
\end{bmatrix}
\end{equation}

For $1\leq i\leq \gamma-1$, $1\leq j\leq \eta-1$, $I(p_{i,j})$ represents a $p\times p$ circulant permutation matrix (CPM) characterized by the value $p_{i,j}$, where $p_{i,j}\in \{0,1,2,\dots,p-1,\infty\}$. If $p_{i,j}=\infty$, $I(\infty)$ is a $p\times p$ zero matrix. If not, for each row index $0\leq r\leq p-1$, the $(r,(r+p_{i,j}) \mod p)$-entry of $I(p_{i,j})$ is '1', and '0' elsewhere. A QC-LDPC code is called fully connected, if there is no $I(\infty)$ in the parity check matrix $H$. The necessary and sufficient condition for the existence of a cycle with length 2k in the Tanner graph is as follows \cite{fossorier2004quasicyclic}:
\begin{equation}
    \sum_{i=0}^{k-1}(p_{m_i,n_i}-p_{m_i,n_{i+1}})\equiv 0 \mod p,\label{eq2}
\end{equation}
where $n_k = n_0, m_i\neq m_{i+1}, n_i\neq n_{i+1}$ for all $0\leq i \leq k-1$. 

A path of length $k$ denoted by $P_k$ is a sequence of nodes $v_0v_1v_2...v_k$ in $G$, where $v_0, v_1, ..., v_k$ are all distinct, $(v_i,v_{i+1})\in E$ for all $0\leq i \leq k-1$. The distance $d(u,v)$ between two vertices $u$ and $v$ is the length of the shortest path between $u$ and $v$, and the diameter of a graph $G$ is defined as $diam(G)=max\{d(u,v)|u,v\in V(G)\}$. A cycle of length $k$ denoted by $C_k$ is a closed path which means $v_0=v_k$ and the other vertices are all distinct. Girth $g$ is the length of the minimum cycle in Tanner graph, which plays an important role in the performance of LDPC codes.

For a subset $S\subseteq L$, we denote the neighborhoods of $S$ as $N(S)$, then the induced subgraph generated by $S$ is defined as $G[S]=(S \cup N(S),E')$, where $E'=\{(v,c)\in E(G)|v \in S, c\in N(S)\}$. An $(a,b)$ trapping set is an induced subgraph generated by a subset $S$ with $|S|=a$ and there are $b$ vertices of odd degree in $N(S)$. Moreover, elementary trapping sets (ETSs) are special trapping sets whose all check nodes are of degree 1 or 2.

For a given ETS, a \emph{variable node (VN)} graph $G_{VN}=(V_{VN},E_{VN})$ \cite{koetter2004pseudo} is constructed by removing all degree-1 check nodes, defining variable nodes of the ETS as its vertices and degree-2 check nodes connecting the variable nodes as its edges.

If the Tanner graph is variable-regular, assume $d_L(v)=\gamma$ for all $v\in L$, then there is a one-to-one correspondence between an $(a,b)$-ETS and its VN graph. By the definition of VN graph, we can get it from the $(a,b)$-ETS; conversely, given a VN graph, we can add a check node on each edge and add $\gamma-d_{G_{VN}}(u)$ degree-1 check nodes for all $u\in V_{VN}$ with $d_{G_{VN}}(u)<\gamma$ to get the $(a,b)$-ETS. 

For $G_{VN}=(V_{VN},E_{VN})$ from an $(a,b)$-ETS with $d_L(v)=\gamma$, we have $|V_{VN}|=a,|E_{VN}|=\frac{1}{2}(a\gamma-b).$ Obviously, when $\gamma$ is odd, $a$ and $b$ must have the same parity; when $\gamma$ is even, $b$ must be even, too. Also, if the girth of Tanner graph is more than 4, the VN graph of its ETS must be a simple graph.

\begin{definition}\label{def1}
A theta graph, denoted by $\theta(a,b,c)$ is formed by three internally disjoint paths with the same pair of endpoints, of length respectively a,b,c, where $a\leq b\leq c$ and $b\geq 2$.
\end{definition}

We use the following notation in extremal graph theory to deduce bounds for the size of ETSs:

\begin{definition}\label{def2}
Let $\mathcal{H}$ is a family of graphs, the Tur\text{\'a}n number $ex(n,\mathcal{H})$ is the maximum number of edges in any graph of $n$ vertices that does not contain any subgraph isomorphic to any of $\mathcal{H}$.
\end{definition}

\begin{definition}\label{def3}
A graph $G=(V,E)$ is called the extremal graph of $\mathcal{H}$ if it does not contain any subgraph isomorphic to any of the graphs in $\mathcal{H}$ and $|V|=n, |E|=ex(n,\mathcal{H}).$
\end{definition}

\section{Tur{\textbf{\'a}}n number of theta graph}

\begin{figure}[!t]
    \centering
    \includegraphics[width=2.5in]{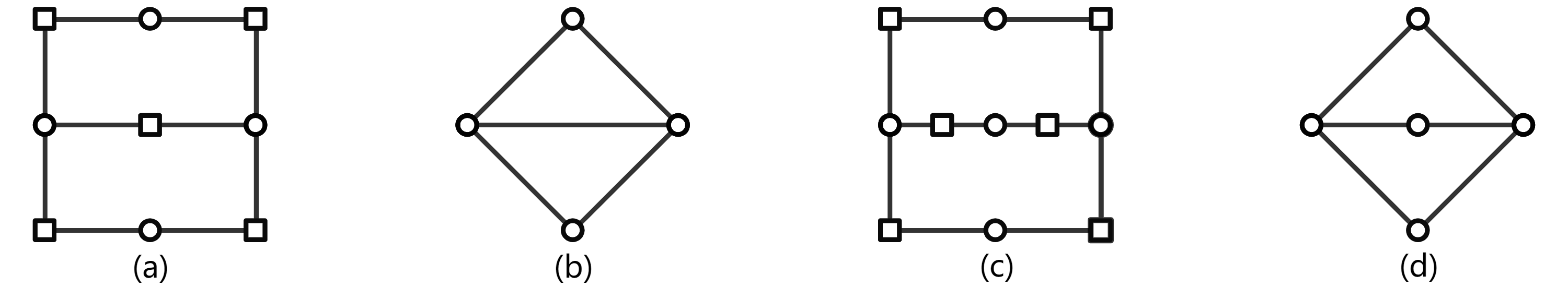}
    \caption{Figures (a) and (b) are two 6-cycles sharing one common check node and its VN graph $\theta(1,2,2)$, respectively. Figures (c) and (d) are two 8-cycles sharing two common check nodes and its VN graph $\theta(2,2,2)$, respectively. In these figures, circles are variable nodes and squares are check nodes.}
    \label{fig1}
\end{figure}

In this section, we derive the Tur\text{\'a}n number for special theta graph, more accurately, theta graph $\theta(1,2,2)$ (Figure \ref{fig1} (b)) for any simple graph and $\theta(2,2,2)$ (Figure \ref{fig1} (d)) for the simple graph with girth $g\geq 4$.

\begin{theorem}[The Tur\text{\'a}n number of $\theta(1,2,2)$]\label{thm4}
For all $n\geq4$, $ex(n,\theta(1,2,2))=\lfloor \frac{n^2}{4}\rfloor.$
\end{theorem}

The Theorem~\ref{thm4} enhances the following result for the case $q=r=2$.

\begin{theorem}[\cite{zhai2021turan}]\label{thm5}
Let $q,r\geq2$ be two integers such that $qr$ is even. Let $k=q+r$ and $n\geq9k^2-3k$. Then
$
ex(n,\theta(1,q,r))=\lfloor \frac{n^2}{4}\rfloor.
$ 
\end{theorem}

\begin{IEEEproof}[Proof of theorem~\ref{thm4}]
By the well-known Tur\text{\'a}n Theorem \cite{turan1941external}, $ex(n,K_{r+1})=\lfloor \frac{(r-1)n^2}{2r}\rfloor$. Then $ex(n,C_3)=\lfloor \frac{n^2}{4}\rfloor$. As the theta graph $\theta(1,2,2)$ contains $C_3$ as its subgraph, so
$
\lfloor \frac{n^2}{4}\rfloor=ex(n,C_3)\leq ex(n,\theta(1,2,2)).
$
Next, we prove $ex(n,\theta(1,2,2))\leq \lfloor \frac{n^2}{4}\rfloor.$ We prove that for any graph $G=(V,E)$ with $|V|=n\geq 4$, if $|E|= \lfloor \frac{n^2}{4}\rfloor+1$, then there must be a theta graph $\theta(1,2,2)$ in $G$ by induction.
\\For $n=4$, $|E|= \lfloor \frac{n^2}{4}\rfloor+1=5$, then $G=K_4-{e}$ is exactly the theta graph $\theta(1,2,2)$. Suppose that the assumption holds for $n=k-1$. 
\\For $n=k$: 
\begin{enumerate}
\item[(i)]If $k$ is odd, assume $k=2m+1$ with $m\geq2$, then $|E|= \lfloor \frac{k^2}{4}\rfloor+1=m^2+m+1$. If every vertex has degree $d_G(v)\geq m+1$, then $|E|\geq \frac{1}{2}(2m+1)(m+1)=m^2+\frac{3}{2}m+\frac{1}{2}>m^2+m+1,$ which contradicts with the assumption $|E|=m^2+m+1$. So, there is a vertex $v_0\in V$, such that $d_G(v_0)\leq m$. Now consider $G-v_0$ with $|V(G-v_0)|=2m,$ and $|E(G-v_0)|\geq m^2+1=\lfloor \frac{(2m)^2}{4}\rfloor+1,$ by the assumption, there is a $\theta(1,2,2)$ in $G-v_0$, which means there is a $\theta(1,2,2)$ in $G$.
\item[(ii)]If $k$ is even, assume $k=2m$ with $m\geq2$, then $|E|= \lfloor \frac{k^2}{4}\rfloor+1=m^2+1$. For the same reason, there is a vertex $v_0\in V$, such that $d_G(v_0)\leq m$. Similarly, consider $G-v_0$ with $|V(G-v_0)|=2m-1,$ and $|E(G-v_0)|\geq m^2-m+1=\lfloor \frac{(2m-1)^2}{4}\rfloor+1,$ by the assumption, there is a $\theta(1,2,2)$ in $G-v_0$, which means there is a $\theta(1,2,2)$ in $G$.
\end{enumerate}
\end{IEEEproof}

The exact value of Tur{\text{\'a}}n number of bipartite graphs is hard to determine \cite{furedi2013history}, we derive an upper bound for theta graph $\theta(2,2,2)$ when there is no $C_3$.

\begin{theorem}[The Tur{\text{\'a}}n number of $\theta(2,2,2)$ and $C_3$]\label{thm6}
For all $n\geq 1$, $ex(n,\{C_3,\theta(2,2,2)\})\leq \frac{n(\sqrt{8n-7}-1)}{4}.$
\end{theorem}

\begin{IEEEproof}
Assume that $G=(V,E)$ is an extremal graph with no $\theta(2,2,2)$ or $C_3$ and $|V|=n.$ If there is a pair of vertices $(x,y)$ with $d(x,y)\geq 4$, we can add the edge $(x,y)$ to $E(G)$ with no new $C_3$ or $C_4$ generated. Therefore, $G+\{(x,y)\}$ still has no $\theta(2,2,2)$ or $C_3$ with one more edge than $G$, which contradicts with the assumption that $G$ is an extremal graph, so the diameter of $G$ is at most 3.\\
We denote $e$ as the number of edges in $G$, $\overline{d}=\frac{1}{n}\sum_{v\in V}d_G(v)$ as the average degree of $G$, and let $\sigma^2=\sum_{v\in V}[\overline{d}-d_G(v)]^2$. We let $D_i=D_i(G)$ be the number of unordered pairs of vertices of $G$ of distance $i$ apart.\\
As the diameter of $G$ is at most 3, we get $\binom{n}{2}=D_1+D_2+D_3=e+D_2+D_3.$ 
As there is no $C_3$ or $\theta(2,2,2)$ in $G$, for a pair of vertices $(x,y)$ with $d(x,y)=2$, the number of their common neighbors is less than or equal to 2. So every pair of vertices of distance 2 apart is counted at most twice in $\sum_{v\in V} \binom{d_G(v)}{2}$, which means $D_2 \geq \frac{1}{2}\sum_{v\in V} \binom{d_G(v)}{2}.$
As $\sum_{v\in V} \binom{d_G(v)}{2}=\frac{1}{2}\sum_{v\in V}d_G(v)^2-e$, $\sum_{v\in V}d_G(v)^2=\sigma^2-\sum_{v\in V}\overline{d}^2+2\sum_{v\in V}\overline{d} d_G(v)=\sigma^2+\frac{4e^2}{n},$
so $D_2\geq \frac{\sigma^2}{4}+\frac{e^2}{n}-\frac{e}{2}.$\\
Then $\binom{n}{2}=e+D_2+D_3\geq \frac{\sigma^2}{4}+\frac{e^2}{n}+\frac{1}{2}e+D_3.$ Finally, we get 
    $$e\leq \frac{\sqrt{8n^3-7n^2-4n\sigma^2-16nD_3}-n}{4}\leq \frac{n(\sqrt{8n-7}-1)}{4}.$$
\end{IEEEproof}

\section{Application to coding theory}
As the number of non-isomorphic structures of ETSs with different values of $a$ and $b$ is hard to calculate and enumerate, in this section, we consider the influence of removing two 6-cycles sharing one common check node (Figure \ref{fig1} (a)) when girth $g=6$ and two 8-cycles sharing two common check nodes (Figure \ref{fig1} (c)) when girth $g=8$ in Tanner graph.

\subsection{Girth $g=6$}

When the girth is 6, if any two 6-cycles in Tanner graph share no common check node (which is called chordless 8-cycles in \cite{amirzade2022protograph}), then there is no theta graph $\theta(1,2,2)$ in VN graph of any $(a,b)$-ETSs. By avoiding this special theta graph, we can get the following theorem:

\begin{theorem}\label{thm7}
For an $(a,b)$-ETS in a Tanner graph with girth 6 and variable-regular degree $d_L(v)=\gamma$, if any two 6-cycles share no common check node, then $b\geq a\gamma-\frac{1}{2}a^2.$
\end{theorem}

\begin{IEEEproof}
For the VN graph $G_{VN}=(V_{VN},E_{VN})$ of an $(a,b)$-ETS with variable-regular degree $\gamma$, we have $|V_{VN}|=a$, $|E_{VN}|=\frac{1}{2}(a\gamma-b)$. As there is no $\theta(1,2,2)$ in $G_{VN}$, we get $|E_{VN}|\leq ex(a,\theta(1,2,2)).$
\\In Theorem~\ref{thm4}, we have proved that $ex(a,\theta(1,2,2))=\lfloor \frac{a^2}{4}\rfloor$ for all $a\geq4$, then we get
$
|E_{VN}|=\frac{1}{2}(a\gamma-b)\leq \lfloor \frac{a^2}{4}\rfloor,
$
which implies
$
b\geq a\gamma-\frac{1}{2}a^2.
$
\end{IEEEproof}

We notice that this bound is tighter than the bound $b\geq a\gamma-\frac{2a^3}{4a-3}$ in \cite{amirzade2022protograph}, as we get the exact value of $ex(n,\theta(1,2,2))$, which means the $(a,b)$-ETS that does not satisfy the bound can not exist in Tanner graph. 

By the famous Tur\text{\'a}n Theorem \cite{turan1941external}, $ex(n,C_3)=\lfloor \frac{n^2}{4}\rfloor=ex(n,\theta(1,2,2)),$ we notice that: for fixed $b$, lifting the girth from 6 to 8 has the same effect as requiring any two 6-cycles sharing no common check nodes on eliminating small ETSs. This observation prompts us to consider the situation of girth $g=8$ in the next subsection.

\begin{table}
\begin{center}
\caption{The minimum size of an $(a,b)$-ETS, $b<a$, in an LDPC code with girth 6, $\gamma=3$ and any two 6-cycles sharing no common check nodes}
\label{tab1}
\begin{tabular}{|c|c|c|}
\hline
b & The minimum $a$ satisfying $b\geq a\gamma-\frac{a^2}{2}$ & The actual $a$  \\ \hline
0 & 6                      & 6                       \\ \hline
1 & 7                      & 7                       \\ \hline
2 & 6                      & 6                       \\ \hline
3 & 5                      & 5                       \\ \hline
\end{tabular}
\end{center}
\end{table}

\begin{table}
\begin{center}
\caption{The minimum size of an $(a,b)$-ETS, $b<a$, in an LDPC code with girth 6, $\gamma=4$ and any two 6-cycles sharing no common check nodes}
\label{tab2}
\begin{tabular}{|c|c|c|}
\hline
b & The minimum $a$ satisfying $b\geq a\gamma-\frac{a^2}{2}$ & The actual $a$  \\ \hline
0 & 8                      & 8                       \\ \hline
2 & 8                      & 8                       \\ \hline
4 & 7                      & 7                       \\ \hline
\end{tabular}
\end{center}
\end{table}

Table~\ref{tab1} and Table~\ref{tab2} list the minimum $a$ calculated by the bound $b\geq a\gamma-\frac{a^2}{2}$ for the case $\gamma=3$ and 4, respectively, which is actually the size of an $(a,b)$-ETS that can exist under the condition we set. Also, these values are coincide with the results obtained by Amirzade \cite{amirzade2022protograph} using computer to enumerate. Obviously, the complexity of calculating the minimum $a$ by this bound is significantly lower than that through enumeration by computer.

\subsection{Girth $g=8$}

When the girth is 8, we set a restriction on Tanner graph that any two 8-cycles share at most one common check nodes, which means no $\theta(2,2,2)$ in VN graph.

\begin{theorem}\label{thm8}
For an $(a,b)$-ETS in a Tanner graph with girth 8 and variable-regular degree $d_L(v)=\gamma$, if any two 8-cycles share at most one common check nodes, then $b\geq a\gamma-\frac{a(\sqrt{8a-7}-1)}{2}.$
If $b<a$,then $a>\frac{1}{2}(\gamma^2-\gamma+2).$
\end{theorem}

\begin{IEEEproof}
By the conditions, there is no $\theta(2,2,2)$ or $C_3$ in VN graph $G_{VN}=(V_{VN},E_{VN})$ of any ETS in Tanner graph. According to Theorem~\ref{thm6}, we get
$
|E_{VN}|\leq ex(a,\{C_3,\theta(2,2,2)\})\leq \frac{a(\sqrt{8a-7}-1)}{4},
$
i.e.
$
b\geq a\gamma-\frac{a(\sqrt{8a-7}-1)}{2}.
$\\
If $b<a$, then $a>b\geq a\gamma-\frac{a(\sqrt{8a-7}-1)}{2},$
so we get
$a>\frac{1}{2}(\gamma^2-\gamma+2).$
\end{IEEEproof}

For specific $\gamma$, using this bound can theoretically prove that several small ETSs do not exist. Although sometimes the minimum $a$ obtained by the bound is smaller than the accurate value, using this bound can significantly reduce the complexity of enumeration.

\begin{figure}[!t]
    \centering
    \includegraphics[width=2.5in]{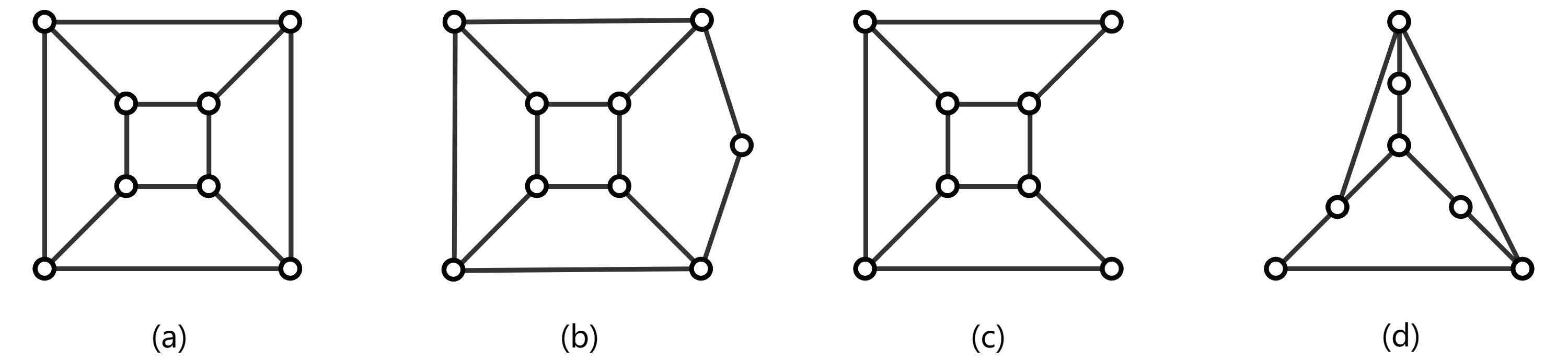}
    \caption{Figures (a), (b), (c) and (d) are the VN graphs of (8,0), (9,1), (8,2) and (7,3)-ETSs with girth 8 and $\gamma=3$, respectively.}
    \label{fig2}
\end{figure}

\begin{corollary}\label{cor9}
For an LDPC codes with girth 8 and $\gamma=3$ with a Tanner graph whose any two 8-cycles sharing at most one common check node, if $b<a$, the smallest sizes of an $(a,b)$-ETS are $a=8,9,8,7$ for $b=0,1,2,3$, respectively.
\end{corollary}

\begin{IEEEproof}
Suppose $b=0$, the minimum $a$ satisfying $b\geq a\gamma-\frac{a(\sqrt{8a-7}-1)}{2}$ is $a=8$.
\\For $b=1$ and 2, although the minimum $a$ satisfying $b\geq a\gamma-\frac{a(\sqrt{8a-7}-1)}{2}$ is $a=7$ and 6, the accurate value is 9 and 8, respectively.
\\For $b=3$, the minimum $a$ satisfying $b\geq a\gamma-\frac{a(\sqrt{8a-7}-1)}{2}$ is $a=7$.
\\Figure \ref{fig2} shows the VN graphs of (8,0), (9,1), (8,2) and (7,3)-ETSs respectively.
\end{IEEEproof}

\begin{corollary}\label{cor10}
For an LDPC codes with girth 8 and $\gamma=4$ with a Tanner graph whose any two 8-cycles sharing at most one common check node, if $b<a$, the smallest sizes of an $(a,b)$-ETS are $a=11,11,10,9$ for $b=0,2,4,6$, respectively.
\end{corollary}

\begin{IEEEproof}
Suppose $b=0$ and 2, the minimum $a$ satisfying $b\geq a\gamma-\frac{a(\sqrt{8a-7}-1)}{2}$ is $a=11$.
\\For $b=4$, the minimum $a$ satisfying $b\geq a\gamma-\frac{a(\sqrt{8a-7}-1)}{2}$ is $a=10$. 
\\For $b=6$, although the minimum $a$ satisfying $b\geq a\gamma-\frac{a(\sqrt{8a-7}-1)}{2}$ is $a=8$, the accurate value is 9.
\end{IEEEproof}

\section{Constructions and Numerical Results}

In this section, we construct QC-LDPC codes with girth 8 and any two 8-cycles sharing at most one common check nodes, which indicates that there is no subgraph shown in Figure \ref{fig1} (c) in the Tanner graph. In order to remove all subgraphs that are isomorphic to Figure \ref{fig1} (c), we check all 8-cycles in the Tanner graph according to the equation (\ref{eq1}), which deduces a necessary and sufficient condition for the existence of a 8-cycle. Indeed, for each pair of variable nodes, we check if there exist three 4-paths with the same lifting value that have the two variable nodes as their endpoints. If so, they will form the graph in Figure \ref{fig1} (c).

\begin{figure}[!t]
    \centering
    \includegraphics[width=2.5in]{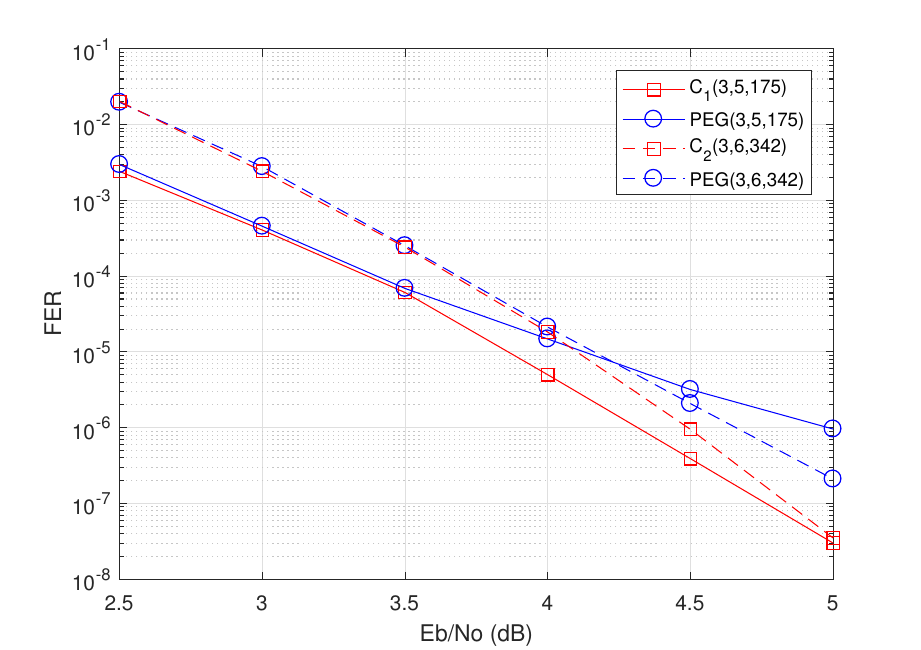}
    \caption{FER performance of $C_1$, $C_2$ and their counterparts}
    \label{fig3}
\end{figure}

By eliminating any two 8-cycles with two common check nodes, we construct the following parity check matrices of (3,5) and (3,6)-regular QC-LDPC codes with girth 8 and lifting degree $p=35, 57$ respectively, which are found by a random search with a cycle controlling process:
\begin{equation}
    H_1 = 
\begin{bmatrix}
    I(0) & I(0) & I(0) & I(0) & I(0)\\
    I(0) & I(4) & I(8) & I(10) & I(21)\\
    I(0) & I(30) & I(15) & I(3) & I(29)\\
\end{bmatrix}
\end{equation}

\begin{equation}
    H_2 = 
\begin{bmatrix}
    I(0) & I(0) & I(0) & I(0) & I(0) & I(0)\\
    I(0) & I(19) & I(10) & I(51) & I(52) & I(26)\\
    I(0) & I(13) & I(56) & I(49) & I(36) & I(27)\\
\end{bmatrix}
\end{equation}

We present simulation results to verify the frame-error-rate (FER) performance in the error floor region of the proposed codes in Figure \ref{fig3}. $C_1$ and $C_2$ are QC-LDPC codes constructed from $H_1$ and $H_2$ with length $N=175$ and 342,respectively, while their counterparts are (3,5) and (3,6)-regular QC-LDPC codes with the same length designed from the progressive-edge-growth (PEG) algorithm. The sum-product algorithm (SPA) is employed to decode the codes over the additive white Gaussian noise (AWGN) channel with binary phase shift keying (BPSK) modulation. Figure \ref{fig3} shows that $C_1$ and $C_2$ have better FER performance in the error floor region, compared with their counterparts.

Note that the method we proposed in this paper (avoiding the presence of two 8-cycles with two common check nodes) is applicable to regular QC-LDPC codes with arbitrary row weights, column weights and whether they are fully connected or not. As we have deduced the Tur{\'a}n number of theta graph $\theta(2,2,2)$ by graphic methods, all $(a,b)$-ETSs that do not satisfy the bound $b\geq a\gamma-\frac{a(\sqrt{8a-7}-1)}{2}$ can be eliminated, as long as the column weight of the regular QC-LDPC code is $\gamma$.

\section{Conclusion}

In this paper, for fixed variable-regular degree $d_L(v)=\gamma$, we consider requiring any two 6-cycles sharing no common check node when the girth is 6, which means there is no theta graph $\theta(1,2,2)$ in VN graph of any $(a,b)$-ETS in Tanner graph. By determining the accurate value of Tur{\text{\'a}}n number for $\theta(1,2,2)$, we prove that all $(a,b)$-ETSs in Tanner graph must satisfy the bound $b\geq a\gamma-\frac{1}{2}a^2$, which improves the bound $b\geq a\gamma-\frac{2a^3}{4a-3}$ in \cite{amirzade2022protograph}. Also, we notice that the minimum $a$ calculated by $b\geq a\gamma-\frac{1}{2}a^2$ is coincide with the values obtained by enumeration. When the girth rises to 8, we deduce an upper bound of $ex(n,\{C_3,\theta(2,2,2)\})$ and prove that if any two 8-cycles share at most one common check node, then all $(a,b)$-ETSs are free of with $(a,b)$ not satisfying the bound $b\geq a\gamma-\frac{a(\sqrt{8a-7}-1)}{2}$. This bound is applicable to regular QC-LDPC codes with arbitrary row weights, column weights and whether they are fully connected or not. The simulation results show that the QC-LDPC codes designed in this paper have good performance with lower error floor.

\bibliography{reference.bib}

\bibliographystyle{IEEEtran}

\end{document}